# Polar coordinates, special relativity and CAS

**Bernhard Rothenstein* and Doru Păunescu***

*Department of Physics,   **Department of Mathematics,
University "Politehnica" of Timişoara

**Abstract**   We investigate the conditions under which computer programs represent correctly, in polar coordinates, the relativistic transformation equations for the space-time coordinates of the same event.

Special relativity works with transformation equations. They establish a relationship between the space-time coordinates of the same event, measured by observers of the inertial reference frames in relative motion **K(XOY)** and **K'(X'O'Y')**. **K** is the stationary reference frame, **K'** is the moving reference and all conditions are met top be in the standard arrangement.

A transformation equation presents in its left side a space coordinate of an event measured in **K**. In its right side it presents the space and the time coordinates of the same event as measured in **K'** and the relative velocity **V** of **K'** relative to **K**. The space coordinates could be Cartesian ($x,y$ in **K** and $x',y'$ in **K'**) or polar ($r,\theta$ in **K** and $r',\theta'$ in **K'**). In many, cases we use transformation equations that present, in both their sides, physical quantities measured in the same reference frame. The computer becomes useful when we establish a relationship between the space-time coordinates of events that take place on a given profile at rest say in **K'** and the space-time coordinates as detected from **K** relative to which the profile moves with constant velocity **V = β•c** (where **β =** 0.8 in our paper).

Consider the photographic detection of the space-time coordinates of a distant event. Let $E'(x' = r'\cos\theta', y' = r'\sin\theta', t')$ be the space-time coordinates of an event that takes place on a profile $P'(x',y') = P'(r',\theta')$ at rest in **K'**. If the profile is luminous and a light signal that leaves the point $M'(r',\theta')$ at a time $-\dfrac{r'}{c}$ being received at the origin **O'** of **K'** at a zero time, a photographic detection of point $M'$ taking place. The events involved in the emission of the light signal $E'(0,0,0)$ associated with its reception at **O'**. Event $E'(0,0,0)$ has the same zero space-time coordinates in all inertial reference frames in relative motion. The Lorentz-Einstein transformations tell us that when detected from **K** event $E'$ has the space-time coordinates $E(x = r\cos\theta, y = r\sin\theta, t = \dfrac{r}{c})$ related by

$$x = \gamma r'(\cos\theta' - \beta) \tag{1}$$
$$y = r'\sin\theta'. \tag{2}$$

From (1) and (2) we obtain with $r = \sqrt{x^2 + y^2}$

$$r = \gamma r'(1 - \beta\cos\theta') \tag{3}$$

Because the angles $\theta$ and $\theta'$ are related by the aberration of light formula

$$\cos\theta' = \frac{\cos\theta + \beta}{1 + \beta\cos\theta} \qquad (4)$$

we can present the right side of (3) as

$$r = r'\frac{\gamma^{-1}}{1 + \beta\cos\theta}. \qquad (5)$$

If the detected profile is the circle

$$r' = R_0 \qquad (6)$$

then relativists consider that (1) and (2) with $r' = R_0$ represent its photographed shape from $\boldsymbol{K}$ in a parametric representation whereas (3) and (4) represent its photographed shape in a polar representation.

All the formulas from above describe some geometrical transformation and a good (easy to use and accurate) visualization tool is needed. There are many such tools, known under the generic name of **Computer Algebra Systems**: huge collections of mathematical algorithms which combine exact symbolic manipulation, approximate numerical calculation and high quality graphical representation. Our paper use **Mathematica 5.2** but **Maple**, **Matworks** or **Mathcad** produces essentially the same results.

The **CAS** starts to work. Making it to represent the circle (6), to perform the parametric representation (1) and (2) and the polar representation (5), it displays on the screen of the computer the same result presented in Figure 1. The command is:

```
In[1]:= β = 0.8; R₀ = 1;    (*  γ = 1/√(1-β²)   *)

In[2]:= ParametricPlot[
        {{R₀ (Cos[θ'] - β)/√(1 - β²), R₀ Sin[θ']}, {R₀ Cos[θ'], R₀ Sin[θ']}},
        {θ', 0, 2 π}, PlotStyle → {Hue[β], GrayLevel[0]},
        PlotRange → {{-3.5, 1.1}, {-1.1, 1.1}}, AspectRatio → 0.46]
```

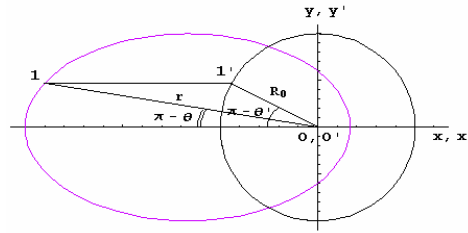

**Figure 1.** Both parametric and polar plots lead to the same result.

To perform polar representation, the standard mathematica kernel is not sufficient and so it calls the graphical package: `` `Graphics` ``. The command is:

```
In[3]:= << Graphics`Graphics`

In[4]:= PolarPlot[{R₀ √(1 - β²)/(1 + β Cos[θ]), R₀}, {θ, 0, 2 π},
        PlotStyle → {Hue[β], GrayLevel[0]},
        PlotRange → {{-3.5, 1.1}, {-1.1, 1.1}}, AspectRatio → 0.46]
```

This comand leads to result shown in Figure 1.

Performing the polar representation of the circle and of (3) it displays on the screen the result shown in Figure 2 that is in total disagreement with the previous results and with the physics of the problem mainly because it does not present horizontal tangents for $y = R_0$ and $y = -R_0$ as required by the relativistic invariance of distances measured perpendicular to the direction of relative motion.

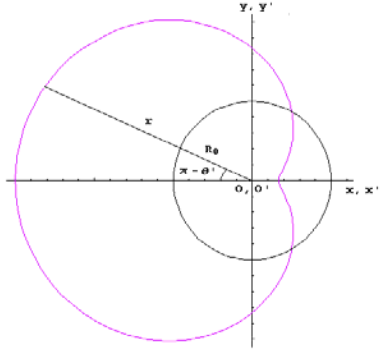

**Figure 2.** Polar plot with wrong parameters

```
In[5]:= PolarPlot[{R0 (1 - β Cos[θ']) / √(1 - β²), R0}, {θ', 0, 2π},
    PlotStyle → {Hue[β], GrayLevel[0]},
    PlotRange → {{-3.5, 1.1}, {-2.1, 2.1}}, AspectRatio → 0.92]
```

It is not a **CAS** error but a human one: **Mathematica** represents in that case $r$ as a function of $\theta'$. Figure (3) we present the way in which $\theta$ depends on $\theta'$.

```
Plot[{ArcCos[(Cos[θ] + β) / (1 + β Cos[θ])], θ}, {θ, 0, π}, AxesLabel → {θ, θ'},
    PlotRange → {{-0.1, π + 0.2}, {-0.1, π + 0.2}}, AspectRatio → 0.86,
    PlotStyle → {GrayLevel[0], Dashing[{0.02, 0.02}]}]
```

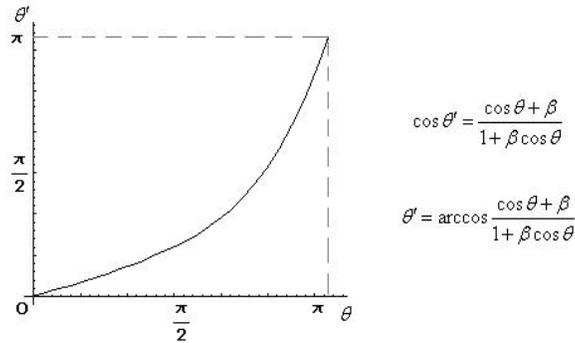

$$\cos\theta' = \frac{\cos\theta + \beta}{1 + \beta\cos\theta}$$

$$\theta' = \arccos\frac{\cos\theta + \beta}{1 + \beta\cos\theta}$$

**Figure 3.** The relationship between the polar angles measured from the involved reference frames.

The pedagogical moral for the beginner is that **CAS** performs polar coordinate representations in accordance with special relativity only in the case when we express the curves as a function of physical quantities measured in the same reference frame, proper physical quantities and the relative velocity of the two frames.

University 'Politehnica' of Timişoara,
Regina Maria Square no.2, 300004 Timişoara, Romania
*E-mail: berhard@etv.utt.ro    **E-mail: dpaunesc@gmail.com